\documentclass[aps,prd,superscriptaddress,showpacs,twocolumn,floatfix]{revtex4}
\usepackage{graphicx}

\newcommand{\ba}{\begin{eqnarray}}
\newcommand{\ea}{\end{eqnarray}}
\newcommand{\beqs}{\begin{eqnarray}}
\newcommand{\eeqs}{\end{eqnarray}}

\begin{document}



\title{Generalized Parton Distributions and Description of
Electromagnetic and Graviton form factors of nucleon.}


\author{O.V. Selyugin \thanks{selugin@theor.jinr.ru} and
  O.V.  Teryaev \thanks{teryaev@theor.jinr.ru}}

\address{\it Bogoliubov Laboratory of Theoretical Physics, \\
Joint Institute for Nuclear Research,
141980 Dubna, Moscow region, Russia \\ }




\begin{abstract}
 The new parametrization of the Generalized Parton Distributions $t$-dependence is proposed.
It allows one to reproduce sufficiently well the electromagnetic form factors of
  the proton and neutron at small and large momentum transfer.
 The description of the data obtained by the Rosenbluth method
  and the polarization method are compared. The results obtained by  the latter
  method are shown to be compatible with the correspondent neutron data.
The impact parameter dependence of  the neutron charge density is examined.
The quark contributions to gravitational form factors of the nucleons are obtained.
\end{abstract}

\maketitle

\section{Introduction}

     The description of the hadron structure is related with our
understanding of the non-perturbative properties of  QCD.
 The essential part of the information about this structure is contained in
 the matrix elements of conserved quark operators.

  The electromagnetic current of a nucleon is
\begin{eqnarray}
 J_{\mu} (P^{'},s^{'}; P,s)  = \bar{u} (P^{'},s^{'}) \Lambda_{\mu} (q,P) u(P,s)  \nonumber \\
 =  \bar{u}(P^{'},s^{'}) (\gamma_{\mu}
 F_{1}(q^2))+\frac{1}{2M} i \sigma_{\mu \nu }q_{\nu }F_{2}(q^{2}))u(P,s),
\end{eqnarray}
where $P,s,  (P^{'},s^{'}) $ are the four-momentum and polarization
of the incoming (outgoing) nucleon
and $q = P^{'}- P $ is the momentum transfer.
The quantity  $  \Lambda_{\mu} (q,P) $ is the nucleon-photon vertex.

The similar expressions for the decomposition of matrix elements
of energy momentum tensors describe the partition
of angular momenta \cite{Ji97} and coupling of quarks and gluons to classical gravity
\cite{Teryaev-s2}.
The latter connection manifests also relation to the equivalence principle
for spin-gravity interactions
\cite{KO,Brodsky00,Silenko:2004ad,Silenko:2006er} and its possible validity
separately for quarks and gluons \cite{Teryaev-s3,Teryaev:2006fk}.

 Two important combinations of the  Dirac and Pauli form factors
are the so-called Sachs form factors \cite{Ernst60}
   \ba
 G^{p}_{E}(t) = F^{p}_{1}(t) + \frac{t}{4M^2} F^{p}_{2}(t);
\ea
 \ba
 G^{p}_{M}(t) = F^{p}_{1}(t)  + F^{p}_{2}(t);
\ea
 with 
  $(t = -q^2 < 0)$.
Their three-dimensional Fourier transform  provides the electric-charge-density
 and the magnetic-current-density distribution \cite{Sachs62}.
Normalization requires
$G^{p}_{E}(0) =1$,  $G^{n}_{E}(0) =0$  corresponding to proton and neutron electric charges;
$G_{M} (0) = (G_{E} (0) +k)  =  \mu $ defines the proton and neutron magnetic moments.
Here  $\mu_p = (1 + 1.79) \frac{e}{2M}$ is the proton magnetic moment and
$k= F_{2}(0)$ is the anomalous magnetic moment: $k_p = 1.79$

  Early experiments at modest $t$, based on the Rosenbluth separation method, suggested
  the scaling behavior of both proton form factors and the neutron magnetic form factor
  approximately described by a dipole form
\ba
G^{p}_{E} \approx \frac{G^{p}_{M}}{\mu_{p}} \approx \frac{G^{n}_{M}}{\mu_{n}}
\approx G_{D} = \frac{\Lambda^4}{(\Lambda^{2} -t)^2},
\ea
which leads to
\ba
 F_{1}^{D} (t) = \frac{4M_{p}^{2} - t \mu_{p}}{4M_{p}^{2} - t }
G_{D};
\ea
\ba
F_{2}^{D} (t) =  \frac{4 k_{p} M_{p}^{2}}{ (4 M^{2}_{p} - t]} G_{D};
\ea
with $\Lambda^2= 0.71$ GeV$^{2}$.
These experiments were based on the Rosenbluth formula \cite{Rosenbluth}
  \ba
\frac{d \sigma}{d \Omega}=\frac{\sigma_{Mott}}{\epsilon (1+\tau)}
 [\tau G^2_{M}(t) + \epsilon G^2_{E}(t)]
 \ea
 where  $\tau= Q^2/4M^2_p$ and $\epsilon = [1+2(1+\tau) \tan^2(\theta_e/2)]^{-1}$
 is a measure of the virtual photon polarization.

   Recently,  better data have been obtained 
 by using of the polarization method \cite{Akhiezer,Arnold}.
 Measuring both transverse
 and longitudinal components of the recoil proton polarization in the electron
 scattering plane,  the data on the ratio
 \ba
 \frac{G^{p}_{E} }{ G^{p}_{M} } = -\frac{ P_t }{ P_l } \frac{ E+E^{'}}{2 M_{p} } \tan(\theta/2)
 \ea
 were obtained. These data manifested a strong deviation from the
 scaling law and, consequently, disagreement with data obtained by the Rosenbluth technique.
 The results consist in   an almost linear decrease of $G^{p}_{E} / G^{p}_{M}$.
 There were attempts to solve the problem by inclusion of additional
 radiative correction terms related to two-photon exchange approximations, for example
 \cite{Guichon}.  
 In recent works \cite{Kuraev1,Kuraev2} the box amplitude is calculated
 when the intermediate state is a proton or the $\Delta$-resonance.
 The results of the numerical estimation show that the present calculation of
 radiative corrections can bring into  better agreement
  the conflicting experimental results
 on proton electromagnetic form factors.
 Note, however, that the precise data of a Rosenbluth measurement of the proton form factors
 at $Q^2$  $ 4.10 \ $ GeV$^2$ \cite{Qat05} lie so high
  that they require very large corrections to move them down to meet the polarization data.

The form factors are related to the first moments
of the Generalized Parton distributions (GPDs) \cite{DMuller94,Ji97,R97,Collins97}.
  Generally,  GPDs depend on the momentum transfer $t$, the average momentum fraction
  $x=0.5 (x_i + x_f)$ of the active quark, and the skewness parameter
  $2 \xi=x_f - x_i$  measures the longitudinal momentum transfer.
  One can choose the special case $\xi=0$ of the nonforward parton densities \cite{R98} ${ \cal{F}}^{a}_{ \xi } (x ; t)$
   for which the emitted and reabsorbed partons carry the same momentum fractions:
\ba
{\cal{H}}^{q} (x,t) \ = \ H^{q}(x,0,t)  \ - \ H^{\bar q} (-x,0,t),
\ea
\ba
{\cal{E}}^{q} (x,t) \ = \  E^{q}(x,0,t) \ - \ E^{\bar q} (-x,0,t).
\ea
The form factors can be represented as moments
\ba
 F_{1}^q (t) = \int^{1}_{0} \ dx  \ {\cal{ H}}^{q} (x, t),
\ea
\ba
 F_{2}^q (t) = \int^{1}_{0} \ dx \  {\cal{E}}^{q} (x,  t),
\ea
following from the sum rules \cite{Ji97,R97}.

Non-forward parton densities also provide information
about the distribution of the parton in impact parameter space \cite{Burk00} which
is connected with $t$-dependence of $GPDs$.
Now we cannot obtain this dependence from the first principles,
but it must be obtained from the phenomenological description with $GPDs$
of the nucleon electromagnetic form factors.

 Choosing a frame where the momentum transfer $r$ is purely transverse $r = r_{\perp} $,
  the two-body contribution to the form factor can be written as \cite{Brodsky81}
\ba
 F^{(2)}(t) =  \int_{0}^{1}   dz  \int_{0}^{1}   \Psi^{*}(x,k_{\perp} +\bar{x} r_{\perp})
  \Psi(x,k_{\perp}) \   \frac{d^{2} k_{\perp}}{ 16 \pi^{3}} 
\ea

As it was shown in \cite{R98,R04}, assuming the Gaussian ansatz
\ba
\Psi(x;k_{\perp}) \sim exp[-\frac{k^2_{\perp}}{2x(1-x)\lambda^2}]
\ea
one can  obtain
\ba
F^{(2)}(q^{2}) = \int^{1}_{0} \ dx \ q^{(2)}(x) e^{(1-x) q^{2}/4x\lambda^2},
\ea
where  $q^{(2)}(x) $ has the meaning of the two-body part of the quark density $q(x)$.
The scale  $\lambda^{2}$ characterizes the average transverse momentum of the  valence quarks in the nucleon and $q^2 =-t$ is momentum transfer.
  This Gaussian ansatz was used in \cite{R98} for describing the form factors of proton.
  However, this ansatz leads to a faster decrease in  $F_1$ at large momentum transfer.
  So there arises an important question about $t$-dependence of  GPDs.

  In  \cite{Goeke01}, it was proposed to use the factorized Regge-like picture
\ba
{\cal{H}}^{q} (x,t) \  \sim  \frac{1}{x^{\alpha' t} }\ q(x).
\ea
 which, however \cite{R04}, does not satisfy some conditions
 in the light-cone representation.

  In \cite{Stol01,Stol02},  two parts of $\Psi(x, k_{\perp})$ were introduced
   which are
 responsible for the interaction at small momentum transfer
   $\Psi(x, k_{\perp})^{soft}$
 and for the interactions at large momentum transfer $\Psi(x, k_{\perp})^{hard}$.
 These functions can reproduce the $t$-dependence of , for example, $F_1$ but at the cost
 of the complicated mechanism and growth of  $F_1$ at  higher value of $t$ .

  In \cite{R04},  the Regge-like picture was used again but now without factorization
\ba
{\cal{H}}^{q} (x,t) \  \sim  \frac{1}{x^{\alpha'  \ (1-x) \ t}} \ q(x).
\ea
 This ansatz reproduces the basic properties of  $F_1$ and provides a good description of
the  ratio of $F_1$ to $F_2$ and, consequently, of the ratio of  $G_E$ to $G_M$.
 A similar approach was examined in \cite{Jenk} with  GPDs form
 $\sim (x/g_0)^{-\alpha(t) (1-x)}$, where $\alpha(t)$ is
 the nonlinear part of the Regge trajectory.

    Note that in \cite{Yuan03,Burk04} it was shown that at  large $x  \rightarrow 1$
    and momentum transfer the behavior of GPDs
  requires a larger power of $(1-x)$ in the  $t$-dependent exponent:
\ba
{\cal{H}}^{q} (x,t) \  \sim  exp [ a \ (1-x)^n \ t ] \ q(x).
\ea
with $n \geq 2$. It was noted that $n=2$ naturally leads to the Drell-Yan-West duality
 between parton distributions at large $x$ and the form factors.

  In other works (see e.g. \cite{Kroll04,Goloskokov:2005sd,Kroll06})
  the description of the $t$-dependence of the GPDs  was developed
  in a some more complicated picture using the polynomial forms:
\ba
{\cal{H}}^{q} (x,t) \  \sim q(x) \  Exp[ f(x) \ t ],
\ea
with
\ba
 f(x) = \alpha' (1-x)^2 ln(\frac{1}{x}) + B_q  (1-x)^2+ A_q  x (1-x)
\ea
or
\ba
f(x) = \alpha' (1-x)^3 ln(\frac{1}{x})+B_q  (1-x)^3 + A_q  x
(1-x)^2
\ea

These parameterizations provide
the good description of the data extracted from \cite{Brash:2001qq}
The
curve from Fig. 6 in \cite{Kroll04} is reproduced in Fig.1.
Note, that
the parametrization of (forward) parton distribution $q(x)$ there
also differs from the other curves at Fig. 1.

 \section{New momentum transfer dependence of GPDs }

Let us modify the original Gaussian ansatz in order to incorporate
the observations of \cite{R98} and \cite{Burk04} and choose
 the $t$-dependence of  GPDs in the form
\ba
{\cal{H}}^{q} (x,t) \  = q(x) \   exp [  a_{+}  \
\frac{(1-x)^2}{x^{m} } \ t ].
\ea
  The value of the parameter $m=0.4$ is fixed by the low $t$  experimental data while
 the free parameters $a_{\pm}$ ($a_{+} $ - for ${\cal{H}}$
and $a_{-} $ - for ${\cal{E}}$) were chosen to reproduce the
experimental data in the whole $t$ region.  Indeed, large $t$ behavior
corresponds to $x \sim 1$ in (10), (11), where the dependence on $m$ is
weak.

The function $q(x)$ was chosen at the same scale $\mu^2=1$ as in \cite{R04},
which is based on the MRST2002 global fit \cite{MRST02}.
 In all our calculations we restrict ourselves, as in other quoted works,  to
 the contributions of $u$ and $d$ quarks.

 Hence, we have
\ba
 u(x) = 0.262 x^{-0.69}(1-x)^{3.50}(1+3.83x^{0.5}+37.65x),
\ea
\ba
 d(x) = 0.061 x^{-0.65}(1-x)^{4.03}(1+49.05x^{0.5}+8.65x).
\ea

Following the standard representation, see for example \cite{R04}, we
have for the Pauli form factor $F_2$ \ba {\cal{E}}^{q} (x,t) \  =
{\cal{E}}^{q} (x) \   exp [  a_{-}  \  \frac{(1-x)^2}{x^{0.4}} \ t
]. \ea
 with
 \ba
{\cal{E}}^{u} (x) \  = \frac{k_u}{N_u} (1-x)^{\kappa_1} \ u(x),
\ea
 \ba
{\cal{E}}^{d} (x) \  = \frac{k_d}{N_d} (1-x)^{\kappa_2} \ d(x),
\ea
 where $\kappa_1 =1.53$ and $\kappa_2=0.31$  \cite{R04}.

According to  the normalization of the Sachs form factors, we have
$$k_u=1.673  , \ \ \  k_d=-2.033,   \  \  \  \   \ N_u=1.53  , \ \ \  N_d=0.946   $$
The parameters $a_{+} = 1.1$ and $a_{-} $
 were chosen to obtain two possible forms of the ratio of the  Pauli and Dirac form factors.
 Below we consider  variant $(I)$ with $a_{-} = 1.1$
 and variant $(II)$ with $a_{-} = 1.4$ which correspond to the description
 of the experimental data obtained by the polarization method
  and the Rosenbluth method, respectively.

 \section{Proton form factors }

 The proton Dirac form factor, multiplied by $t^2$,
  is shown in Fig.1 in comparison
 with  other works and the experimental data.
 It can be seen that the complicated mechanism proposed in \cite{Stol01}
 leads to an increase in these values  at higher $t$.
One can see that our model  reproduces sufficiently well the
behaviour of the experimental data at both high $t$ and low $t$.

\begin{figure}
\includegraphics[width=.4\textwidth]{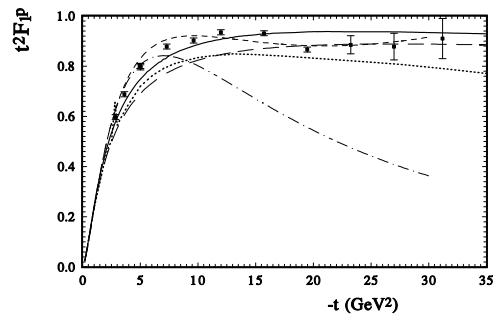} 
\caption{ Proton Dirac form factor multiplied by $t^2$
 (hard line - the present work,   dot-dashed line - \cite{R98};
 long-dashed line - \cite{R04}; dashed line - \cite{Stol02}; dotted line - \cite{Kroll04};
 the  data for $F_1^{p}$ are from \cite{Sill93}
  }
\label{Fig_1}
\end{figure}

The ratio of the Pauli to the Dirac  proton form factors multiplied by $t$
 is shown in Fig.2.
 As it was mentioned above, there are  two different sets of  experimental data.
  Here we are not going to
discuss these two methods and
  different corrections  to them. We just observe that our model describes the results
  of both the  methods by changing the slope of  ${\cal{E} }$ only and present two respective
  variants of this slope $a_{\pm}$.

\begin{figure}
\begin{center}
\includegraphics[width=.4\textwidth]{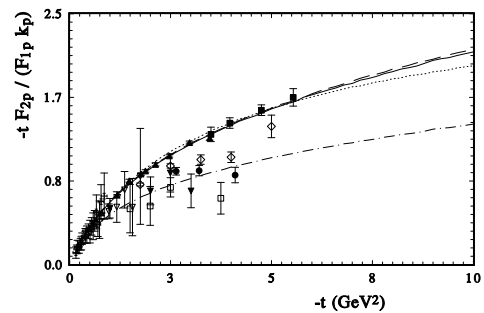}
\end{center}
\caption{Ratio of the Pauli to  Dirac  proton form factor multiplied by $t$
(hard and dot-dashed lines correspond to variant (I) and (II)) of the present  work ,
  dotted line - \cite{Brodsky03}; long-dashed line - \cite{R04})
 ; the data  are from   \cite{Jones00,Gayou01,Gayou02,Punjabi05}.
  }
\label{Fig_2}
\end{figure}

\begin{figure}[!ht]
~\vglue -1.5cm
\includegraphics[width=.4\textwidth]{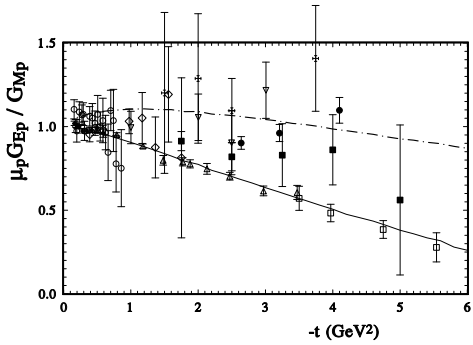}
 \caption{ $\mu_{p} G_{E}^{p}/G_{M}^{p}$ (hard and dot-dashed lines correspond
 to variant (I) and (II)); the experimental data
from  \cite{Gayou01,Gayou02,Punjabi05,Arr05,Hu06,Graw07,Qat05,Qat06}.
 } \label{Fig_3}
\end{figure}
This is supported also by   Fig.3 where $\mu_{p} G_{E}^{p}/G_{M}^{p}$
are shown in comparison with the experimental data.

\section{Neutron form factors }

  Let us now calculate the neutron form factors using the model  developed for the proton.
  The isotopic invariance can be used to relate the proton and neutron GPDs.
 Hence, we do not change any parameter
 and keep the same $t$-dependence of GPDs as in the case of  proton

\ba
{\cal{H}}^{n} (x,t) \  = q_{+}(x)_{n}  \   exp [ 2 a_{+}  \  \frac{(1-x)^2}{x^{0.4} } \ t ],
\ea
where
\ba
q_{+}(x)_{n} = \frac{2}{3} d \ - \frac{1}{3} u.
\ea

 For the Pauli form factors of the neutron we correspondingly have
\ba {\cal{E}}^{n} (x,t) \  =  q_{-}(x)_{n} \   exp [ 2 a_{-}  \
\frac{(1-x)^2}{x^{0.4}}  \ t ], \ea
 with
 \ba
 q_{-} (x)_{n}   =  -\frac{1}{3} \frac{k_u}{N_u} (1-x)^{\kappa_1} \ u(x) + \frac{2}{3}  \frac{k_d}{N_d} (1-x)^{\kappa_2} \ d(x).
\ea
   We take the two values of the slope $a_{-} $
as in the case of the proton form factors, corresponding  to variant
$(I)$ and variant $(II)$ below.

  First, let us calculate $G_{E}^{n}$. The results are shown in Fig. 4.
 Evidently,
 the first variant is in better
 agreement with the experimental data.

\begin{figure}[!ht]
~\vglue -1.5cm
\includegraphics[width=.4\textwidth]{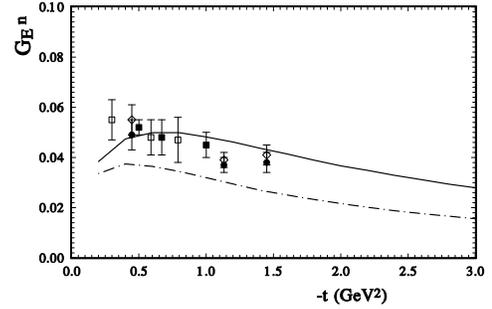}
 \caption{ $G_{E}^{n}$ (hard and dot-dashed lines correspond to variant (I) and (II));
 the experimental data from  \cite{Plaster05,Madey03,Warren04}.
 } \label{Fig_4}
\end{figure}

A more clear situation with our calculations of $G_{M}^{n}$ is shown in Fig.5.
  In this case, it is obvious that the first variant much better describes
 the experimental data, especially at low
  momentum transfer.

\begin{figure}[!ht]
~\vglue -1.5cm
\includegraphics[width=.4\textwidth]{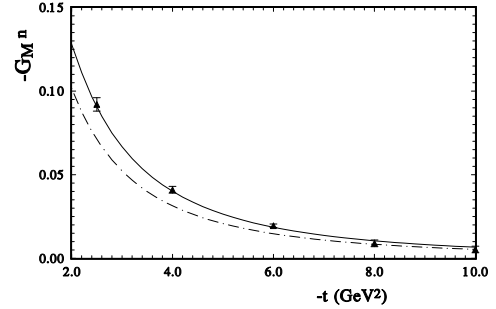}
 \caption{ $G_{M}^{n}$ (hard and dot-dashed lines correspond to variant (I) and (II));
 the experimental data from  \cite{Rock82}.
 } \label{Fig_5}
\end{figure}

 Finally, we present our calculations for the ratio of $G_{E}^{n}/G_{M}^{n}$
 for neutron in Fig. 6.

\begin{figure}[!ht]
~\vglue -1.5cm
\includegraphics[width=.4\textwidth]{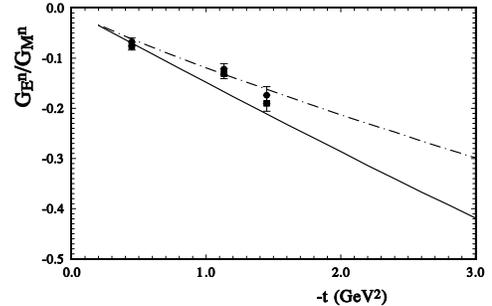}
 \caption{ $G_{E}^{n}/G_{M}^{n}$ (hard and dot-dashed lines corresponds to variant (I)
  and (II)); experimental data from \cite{Plaster05,Madey03}.
 } \label{Fig_6}
\end{figure}

\section{Charge Densities of neutron }

Let us now discuss the neutron structure in the impact parameter
 \cite{SopperT,BurkT,RayT} representation.
we are particularly motivated by recent discussion of the definition of charge density of the  neutron
at small impact parameters corresponding to the  "center" of the neutron \cite{BurkZ,MillerZ}.
     In \cite{MillerZ}, the charge density of the neutron is related to
  $F^n_1(t)$ and calculated using
  phenomenological representation of the $G^n_E(t)$ and $G^n_M(t)$.
\ba
\label{imp}
 \rho(b)& = &\sum_{q} e_{q}  \int \ dx \ q(x,b)  \nonumber \\
 & = & \int d^{2}q F_1 (Q^2 = q^2) e^{i \vec{q} \vec{b} } \\
 & =& \int_{0}^{\infty} \ \frac{q dq}{2 \pi} J_{0}(q b) \
 \frac{G_{E}(q^2) + \tau G_{M}(q^2)}{1+ \tau} \nonumber.
\ea
  $J_{0}$  being a cylindrical Bessel function.
 It differs essentially from the definition of the neutron charge
  distribution in the Breit frame related $G^n_E (t)$.

\ba
 \rho_{G_{E}}(b)
 & = & \int d^{2}q [F_1 (q^2)+ \tau F_{2}(q^2)] \ e^{i \vec{q} \vec{b} } \nonumber \\
 & =& \int_{0}^{\infty} \ \frac{q dq}{2 \pi} J_{0}(q b) \
 G_{E}(q^2).
\ea
\begin{figure}[!ht]
 ~\vglue -1.5cm
\includegraphics[width=.4\textwidth]{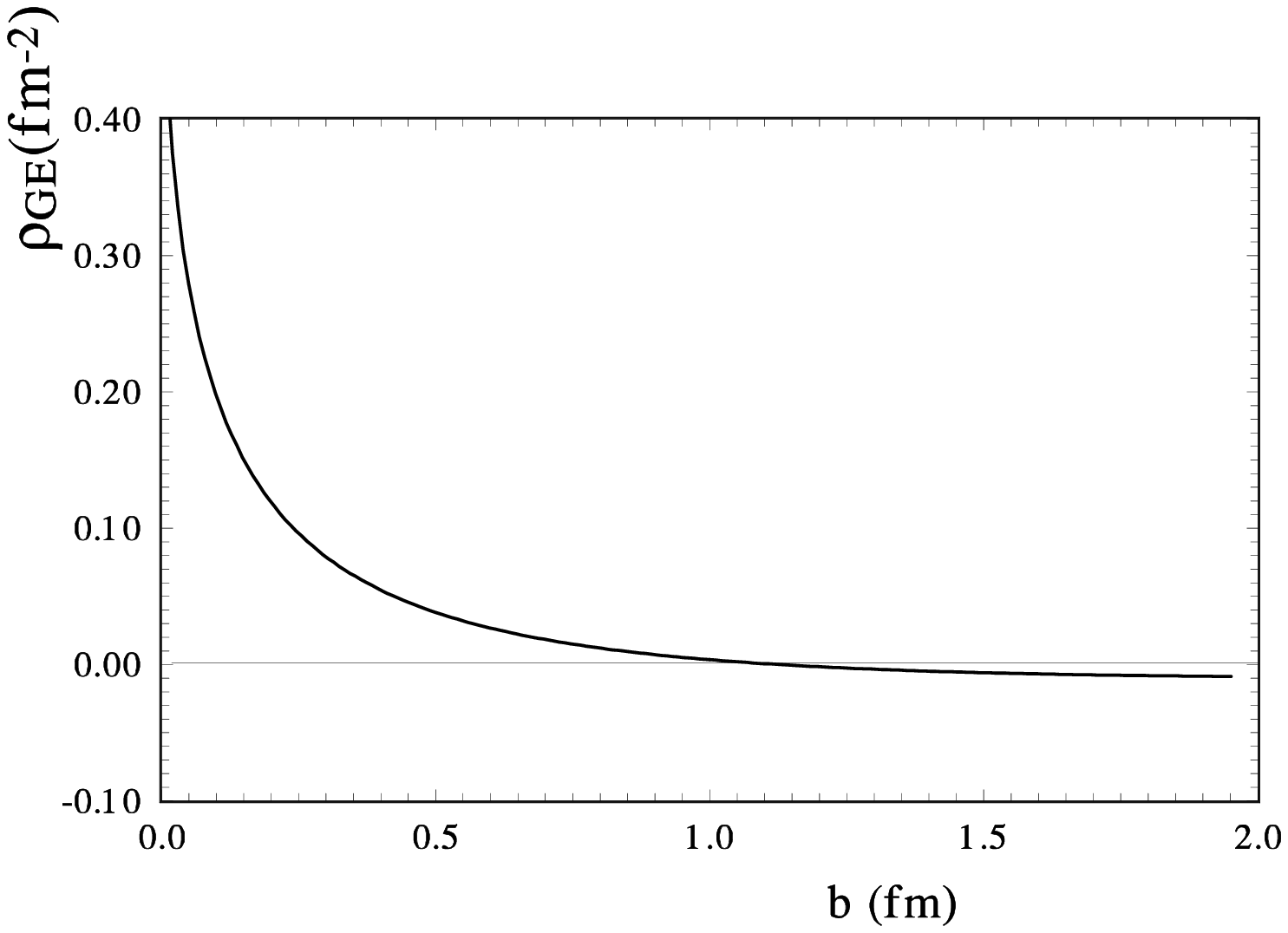}
 \caption{ Parton charge density of the neutron  $\rho_{n}(b)$
  correspound to $G_E(b)$
 } \label{Fig_7}
\end{figure}
\begin{figure}[!ht]
~\vglue -1.5cm
\includegraphics[width=.4\textwidth]{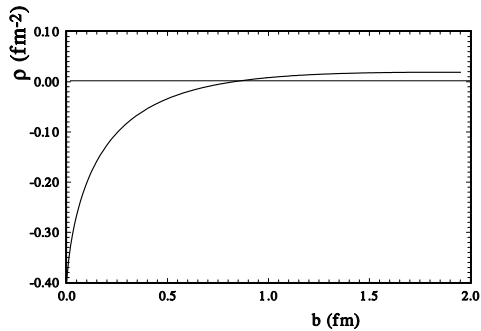}
 \caption{ Parton charge density of the neutron  $\rho_{n}(b)$ correspond to $F_1(b)$
 }.
\label{Fig_8}
\end{figure}

 Using our model of $t$-dependence of  GPDs we may calculate
  both forms of the neutron charge distribution in the impact parameter representation
  and, moreover, determine  separate  contributions of $u$ and $d$ quarks.

The charge distribution $\rho_n(b)$ corresponding to $G^n_E$ is shown in Fig. 7.
  It practically coincides with the result \cite{Arrington-ron}.
 The charge  density of the neutron   $\rho_{n}(b)$ corresponding to $F_1$ is shown in Fig. 8. while the respective
  separate contributions of $u$ and $d$-quarks are shown in Fig. 9.
  We can see that $u$-quarks have the large negative charge density in the centre of the
 neutron.

\begin{figure}[!ht]
\includegraphics[width=.4\textwidth]{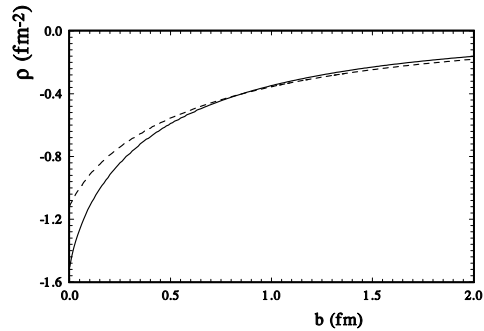}
 \caption{ $u$-quark (hard line) and $(-d)$-quark (dashed line) density of the neutron
 }.
\label{Fig_9}
\end{figure}

\section{Transverse Charge Densities }

\begin{figure}[b]
\hspace{1.5cm}\includegraphics[width=.30\textwidth]{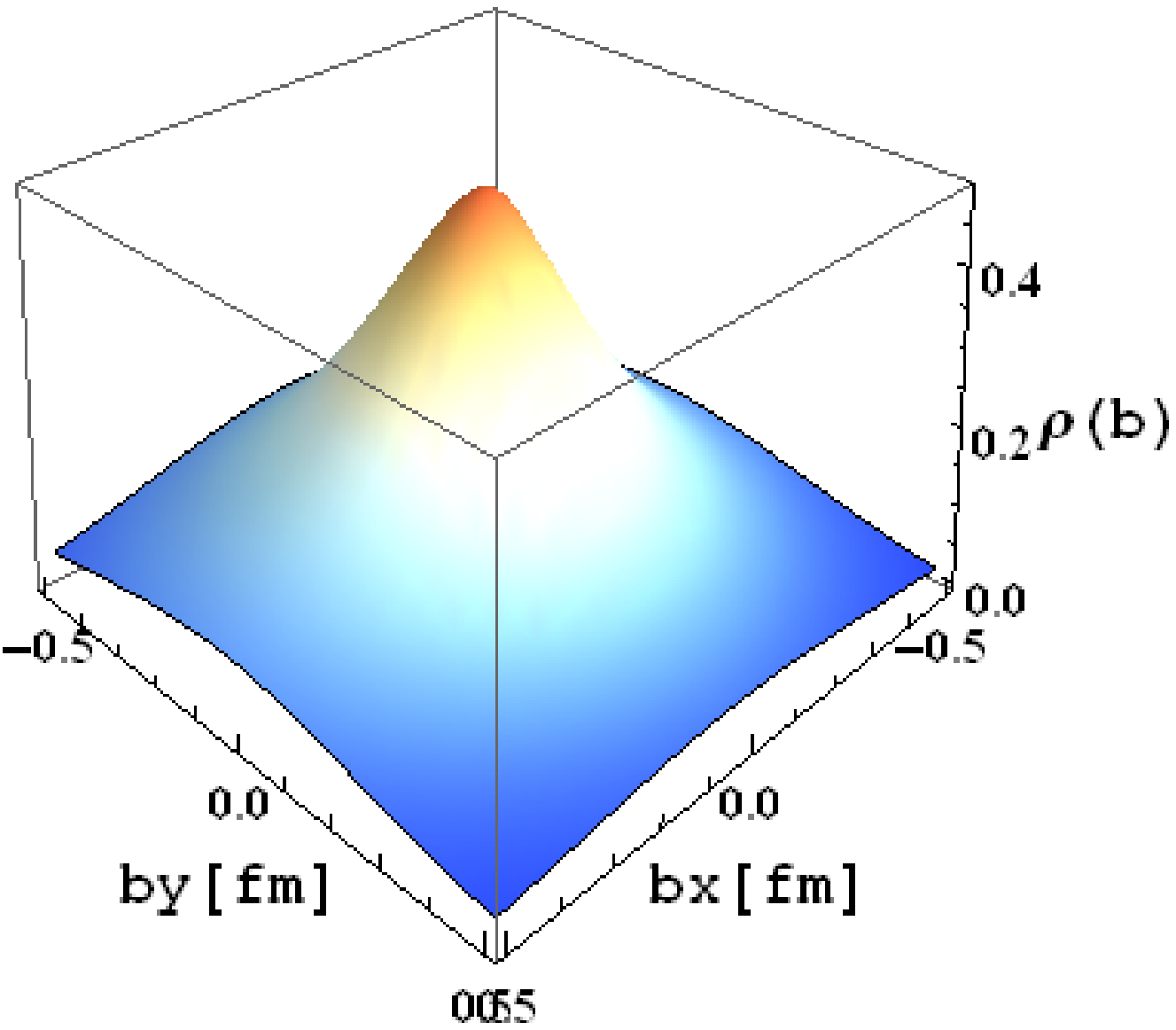}
\includegraphics[width=.25\textwidth]{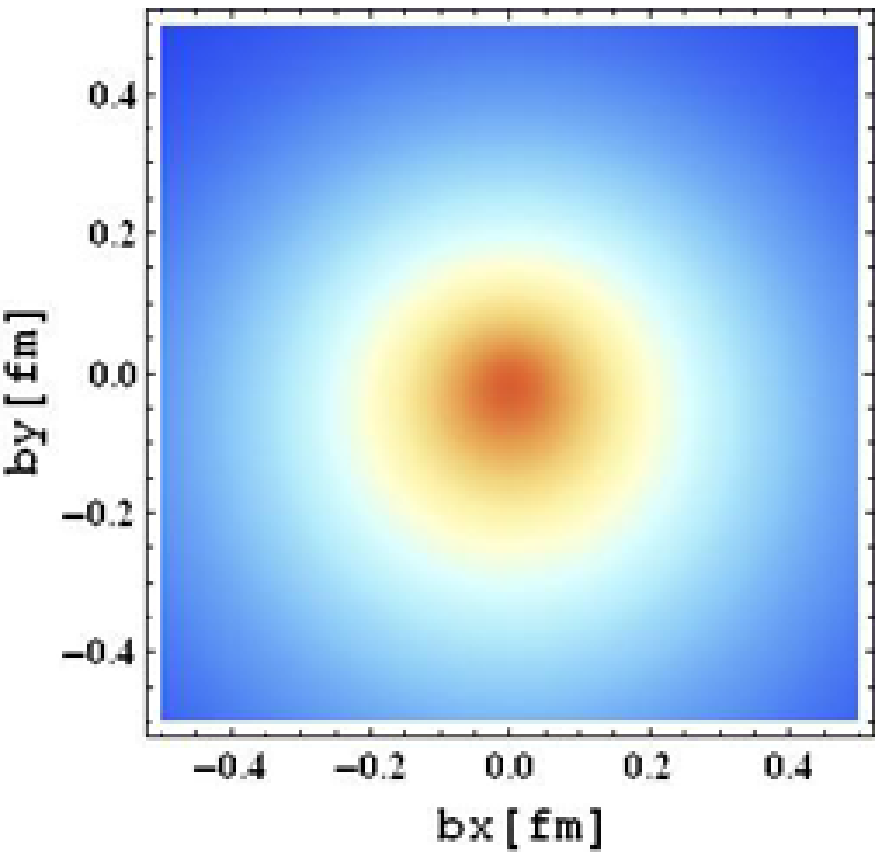}
 \caption{ Quark transverse charge density of the proton eq.(34):
 a)upper panel - back side view; b)low panel -
  view from above.
 }.
\label{Fig_10}
\end{figure}
Recently, the impact parameter distributions were generalized  \cite{Vanderh}
 to the case of spin-flip magnetic
form factor so that in addition to expression (\ref{imp})
one has
\ba
 \rho_{T}^{N}(\vec{b}) =  \rho_{0}^{N}(\vec{b}) + Sin(\phi)  \frac{1}{2 \pi }
\int_{0}^{\infty} dq  \frac{q^2}{2M_{N}} J_{1}(q b) F_{2}(q^2), \label{FT}
\ea
here $\tan(\phi)=b_x/b_y$.
This transverse charge density for proton obtained in the framework of our model is shown
in Figs 10 while the angular-dependent contribution (second term in (\ref{FT})) is shown in Fig. 11.

\begin{figure}[!ht]
\hspace{1.5cm}\includegraphics[width=.30\textwidth]{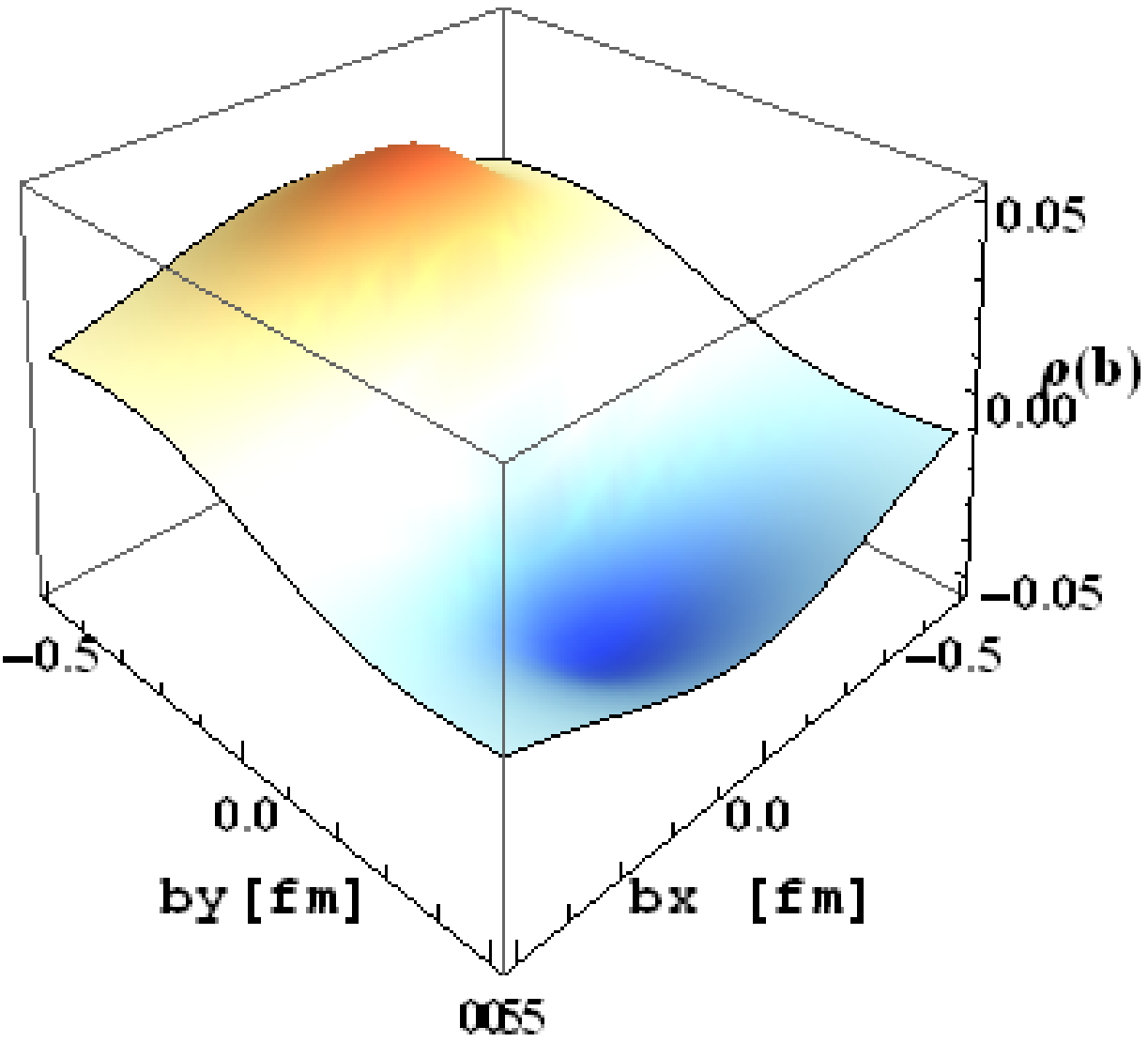}
\includegraphics[width=.25\textwidth]{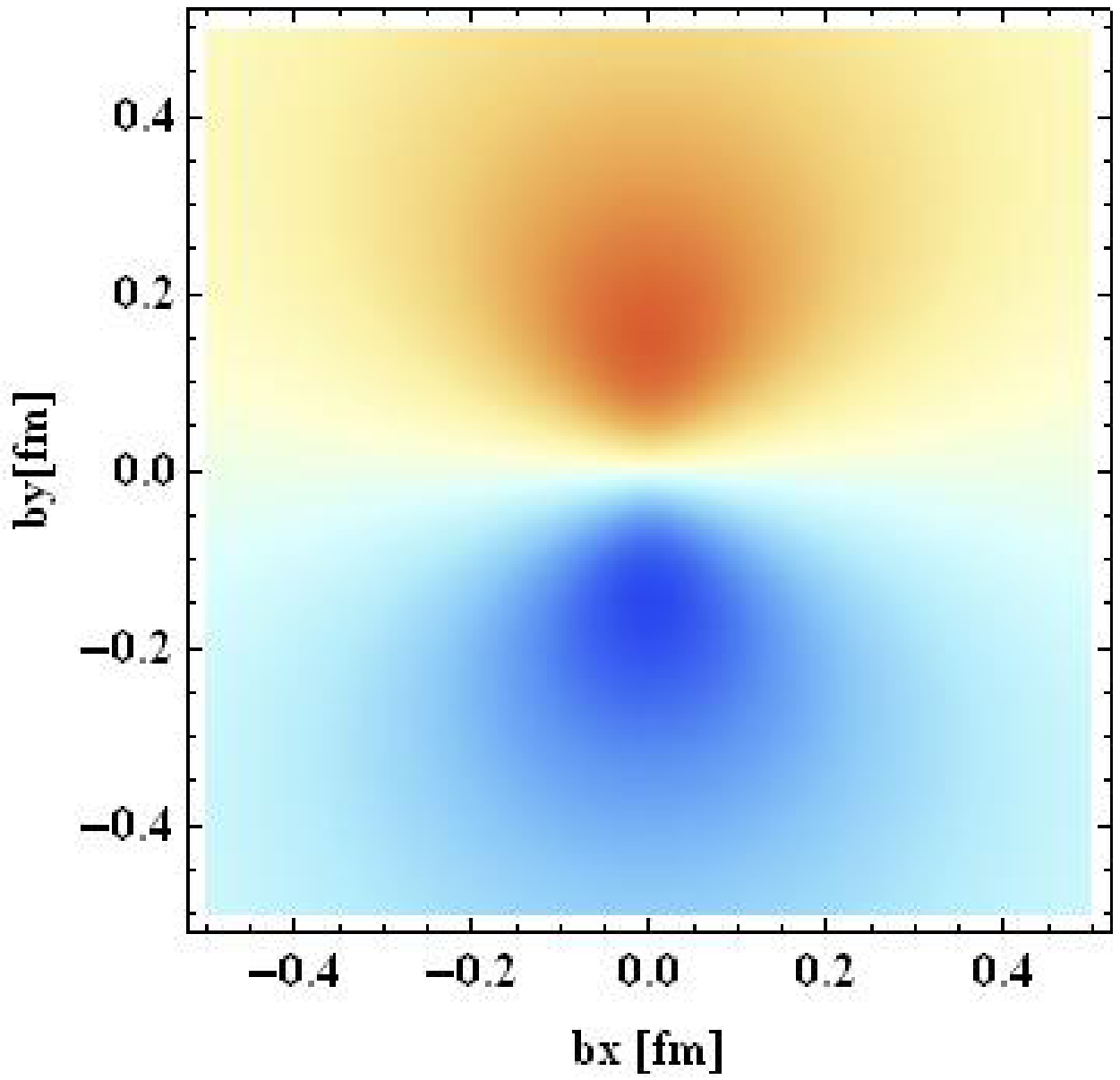}
 \caption{ Angular dependent contribution to quark transverse charge density of the proton  $\rho^{p}_{T}(\vec{b})-\rho^{p}_{0}(\vec{b})$:
  a)upper panel - back side view; b)low panel -
  view from above.}
\label{Fig_11}
\end{figure}

\begin{figure}[t]
\includegraphics[width=.30\textwidth]{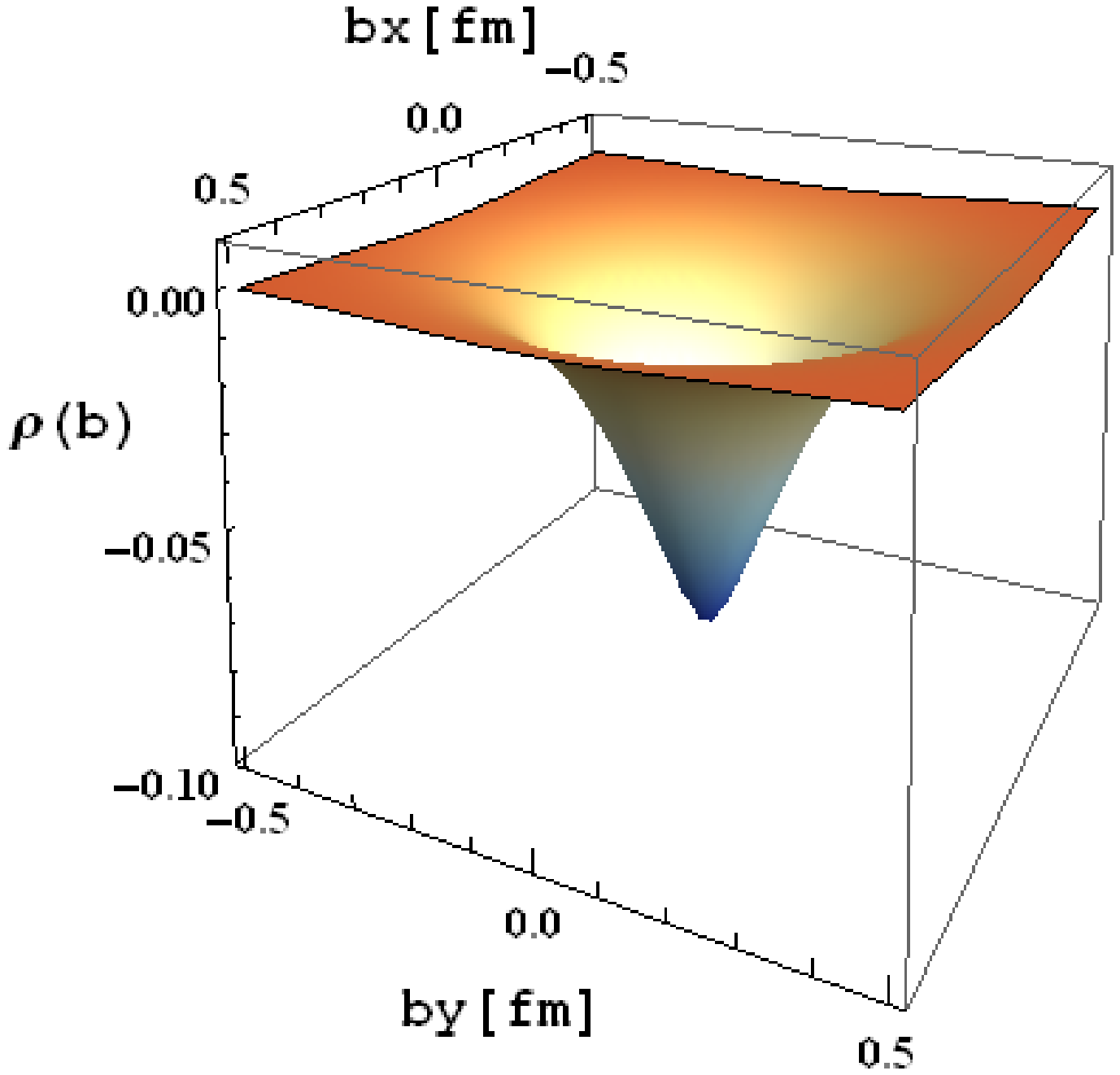}
\includegraphics[width=.25\textwidth]{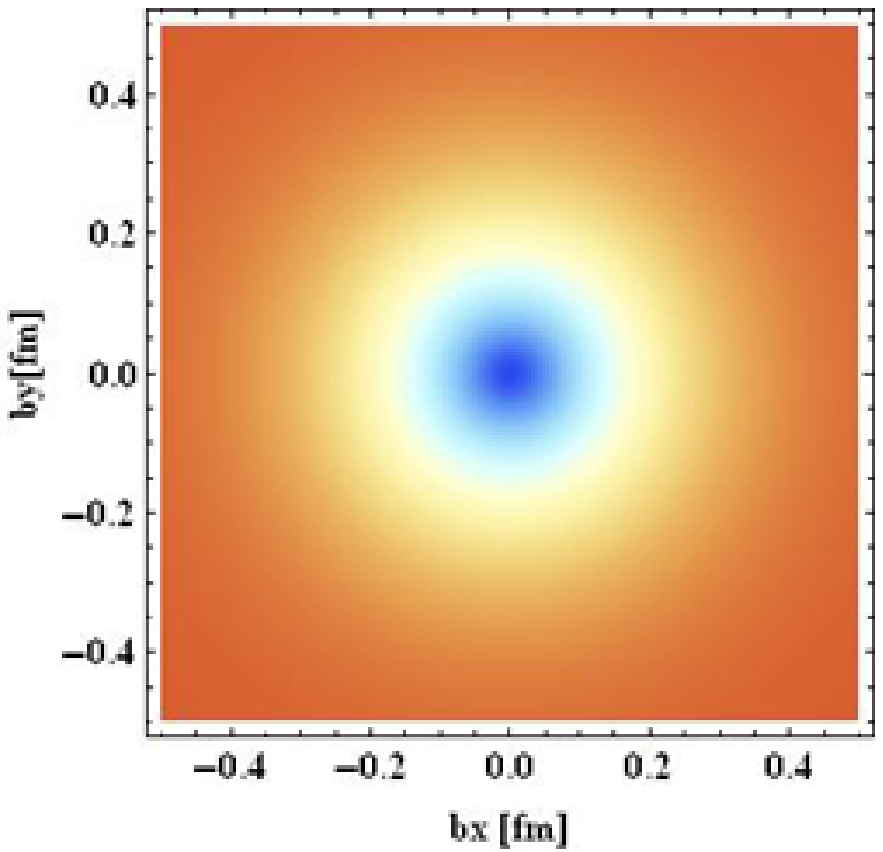}
 \caption{ Quark transverse charge densities of the neutron $\rho^{n}_{0}(\vec{b})$:
a)upper panel -
  with the back side view point;  b)low panel - with the upper view point.
 }.
\label{Fig_12}
\end{figure}

\begin{figure}[!ht]
\includegraphics[width=.30\textwidth]{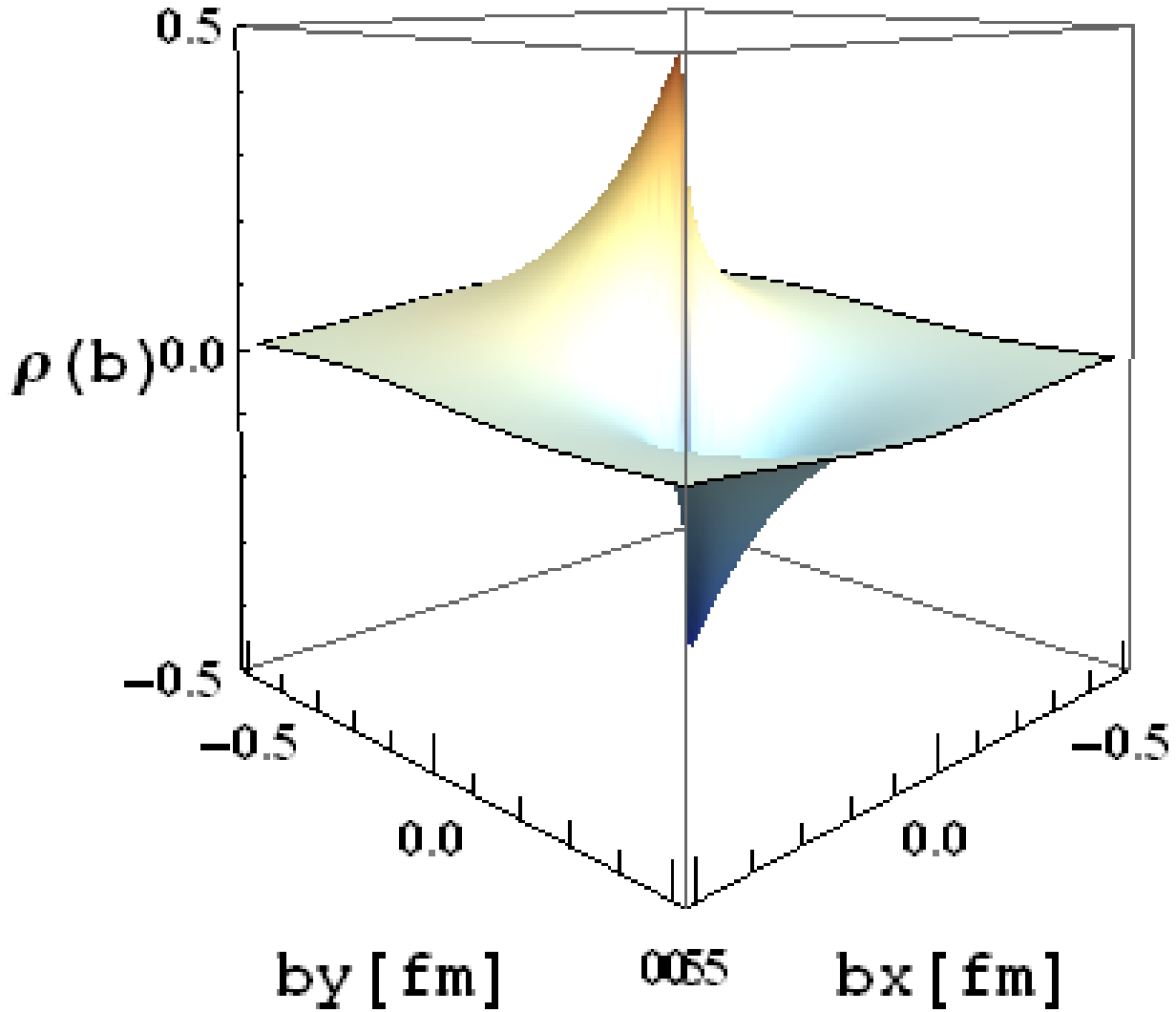}
\includegraphics[width=.25\textwidth]{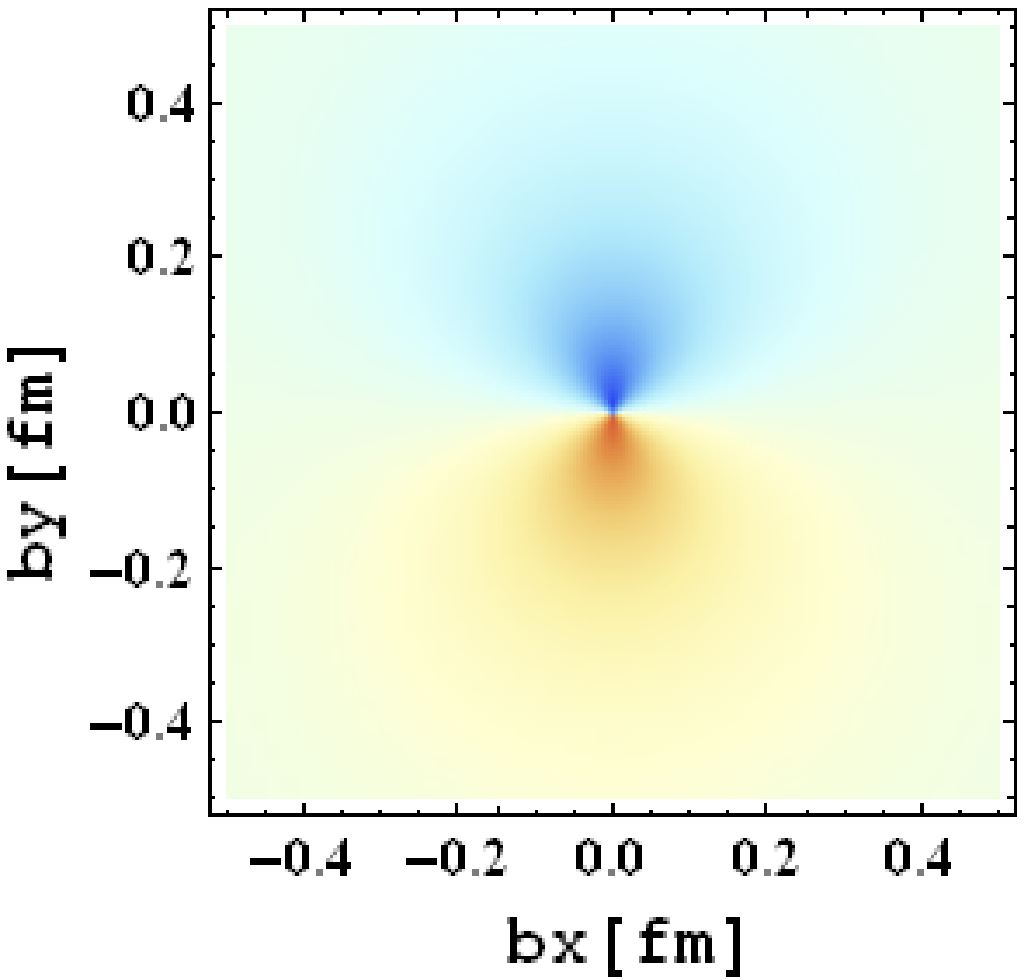}
 \caption{ Quark transverse charge densities of the neutron  $\rho^{N}_{T}(\vec{b})-\rho^{N}_{0}(\vec{b})$:
 a)upper panel -
  with the back side view point;  b)low panel - with the upper view point.
 }
\label{Fig_13}
\end{figure}

The angular dependent part of the density is small and it is practically invisible in Fig.10.
   However, removing the axially symmetric part of the density related to $ \rho_{0}^{N}(\vec{b})$
 one can see that non-symmetric part is reaching the value about $\pm 0.05$ (see Fig.11) at the distance of
 $0.3 \ $fm from the center of the proton, comparable to the size of the valence quark.

   The transverse densities of the neutron presented in Fig. 12 and Fig. 13.
   In this case the symmetric part of the transverse density  $\rho_{0}^{n}(\vec{b})$   is small (compare with Fig. 8).
   while the  non-symmetric part is sufficiently large and concentrated around the center of the neutron.

\section{Gravitational form factors }

   Taking the
 matrix elements of energy-momentum  tensor $T_{\mu \nu}$
 instead of
   the electromagnetic current $J^{\mu}$
  one can obtain the gravitational form  factors of quarks which are related to the second, rather than the first moments of GPDs
\ba
\int^{1}_{-1} \ dx \ x H_q(x,\Delta^2,\xi)  = A_q(\Delta^2); \nonumber \\
\int^{1}_{-1}  \ dx \ x \ E_q (x,\Delta^2,\xi) = B_{q}(\Delta^2).
\ea
 For $\xi=0$ one has
\ba
\int^{1}_{0} dx \ x {\cal{H}}_q(x,t) = A_{q}(t); \,\ \int^{1}_{0}  dx \ x {\cal{E}}_q (x,t) = B_{q}(t).
\ea

\begin{figure}[!ht]
\includegraphics[width=.4\textwidth]{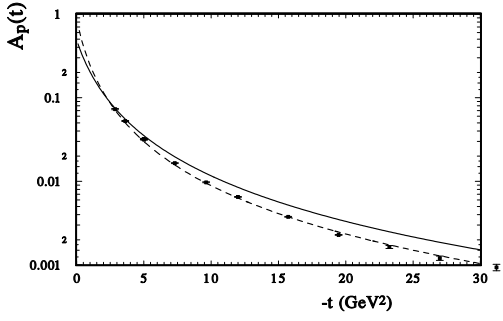}
\caption{ Comparison of gravitational form factor $A_{u+d}$ (hard line)
and  Proton Dirac form factor (dashed line) multiplied by $t^2$,
 the  data for $F_1^{p}$ are from  \cite{Sill93}}.
  \label{Fig_14}
\end{figure}

This representation combined with our model (we use here the first variant of parameters describing
 the experimental data obtained by the polarization method) allows to calculate
 the gravitational form factors of valence quarks and
their contribution (being just their sum) to gravitational form factors of  nucleon.
 Our result for $A_{u+d}(t)$ is shown in Fig.14. Separate  contributions of the $u$ and $d$-
quark distribution are shown in  Fig.15.
 At $t=0$ these contributions equal $A_u(t=0)=0.35$
 and $A_d(t=0)=0.14$. The corresponding calculations for $B_q(t)$ are shown
  in Figs. 16.
  At $t=0$ these contributions equal $B_u(t=0)=0.22$
 and $B_d(t=0)=-0.27$. Hence their sum shows the sort of compensation supporting
 the conjecture \cite{Teryaev-s3,Teryaev:2006fk}about validity of the Equivalence Principle
 separately for quarks and gluons:  $B_{u+d}(t=0)=-0.05$.

%
\begin{figure}[!ht]
\includegraphics[width=.4\textwidth]{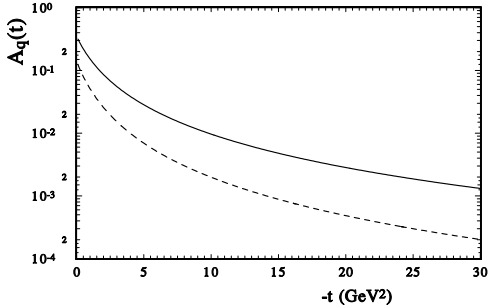}
\caption{ Contributions of the $u$-quark (hard line) and $d$-quark (dashed line)
to the gravitational form factor $A_q$
 }
  \label{Fig_15}
\end{figure}

\begin{figure}[!ht]
~\vglue -1.5cm
\includegraphics[width=.4\textwidth]{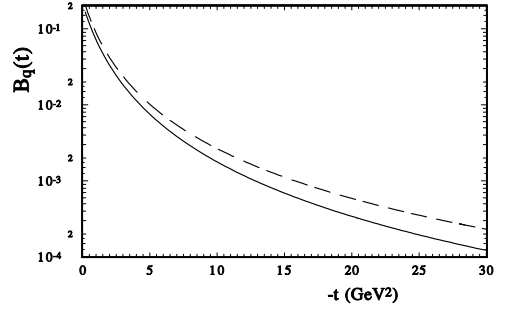}
 \caption{Contributions of the $u$ (hard line) and $d$ (with reversed sign; dashed line) quarks
to the gravitation form factor $B_q$
 } \label{Fig_16}
\end{figure}
Note that nonperturbative analysis within the framework of the lattice OCD indicates
 that the net quark contribution to the anomalous gravitomagnetic moment $B_{u+d}(0)$is close to zero
  \cite{Gockler04,Hagler05}.
\begin{figure}[!ht]
\includegraphics[width=.3\textwidth]{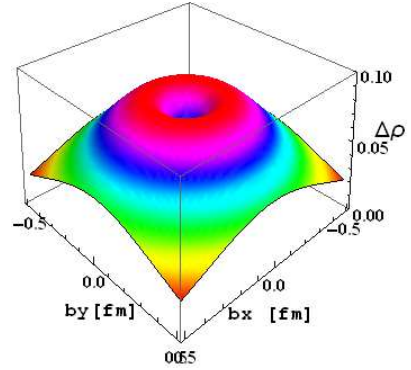}
 \caption{ Difference in the forms of charge density $F^{P}_{1}$
 and  "matter" density ($A$)
 }.
\label{Fig_17}
\end{figure}

\begin{figure}[!ht]
\includegraphics[width=.3\textwidth]{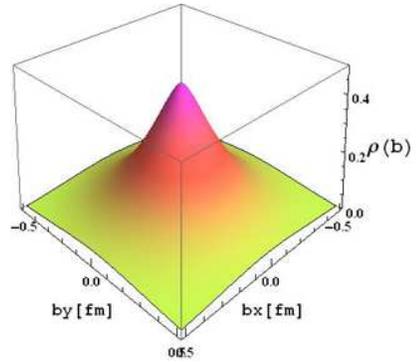}
 \caption{ Densities of gravitational form factor $A$ }
\label{Fig_18}
\end{figure}
Let us compare the distribution of electric charge and matter (that is, gravitational charge) in the nucleon.
For that purpose we generalize  (\ref{FT}) in a straightforward way and introduce the gravitomagnetic transverse density
\ba
 \rho_{T}^{Gr}(\vec{b}) =  \rho_{0}^{Gr}(b) + Sin(\phi) \ \frac{1}{2 \pi }
\int_{0}^{\infty} dq \frac{q^2}{2M_{N}} J_{1}(q b) B(q^2).
\ea
with a matter density
\ba
 \rho_{0}^{Gr}(b) = \frac{1}{2 \pi}
\int^{0}_{\infty}   dq \ q  \ J_{0}(q b) A(q^2).
\ea
In  Fig.17 we compare this matter density with the charge density
(c.f. Section 5) for proton.
The plots for the angular dependent part of transverse density $\rho_{T}^{Gr}(\vec{b})-\rho_{0}^{Gr}(b)$
are shown in fig.18 and fig.19. One can see that it is quite small and its maximal values are
 concentrated almost at the same distances from the center of nucleon as the transverse charge density of the proton.
%
\begin{figure}[!ht]
\includegraphics[width=.35\textwidth]{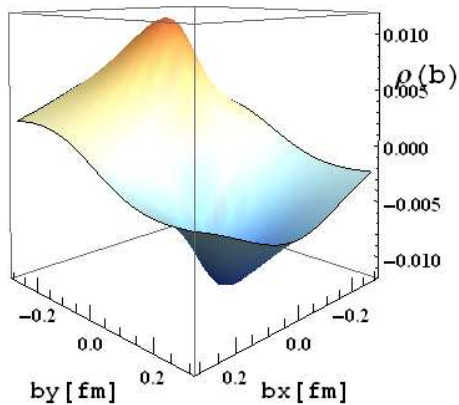}
\caption{ Transverse density $\rho_{T}^{Gr}(\vec{b})-\rho_{0}^{Gr}(b)$
 }
  \label{Fig_19}
\end{figure}

\section{Conclusion }
 We introduced a simple new form of the GPDs
 $t$-dependence based on the Gaussian ansatz corresponding to that
 of the wave function of the hadron. It satisfies the conditions of the non-factorization,
 introduced by Radyushkin, and the Burkhardt condition on the power of $(1-x)^n$
 in the exponential form of the $t$-dependence. With this simple form
  we obtained a good description of the proton electromagnetic form factors.
  Using the isotopic invariance we obtained also a good descriptions of the neutron
  Sachs form factors without changing any  parameters.
    We showed that both sets of the experimental data (obtained by the Rosenbluth
 and polarization methods) on the electromagnetic form factors can be described
  by changing only the slope of the $t$-dependence of the spin-dependent GPD ${\cal{E}}(x,t)$.
   The comparison of these two variants with the neutron form factors gives
  preference to the one which describes the data obtained by the polarization method.

   Our calculations of the charge distribution of the neutron in the impact parameter
   form of $F_1$ coincide with the calculation by Miller obtained from the
   phenomenological forms of $G_E^p$ and $G_{M}^p$. They confirm that respective charge density
   is  negative at small impact parameters.
   On the basis of our results we calculated the contribution of the $u$ and $d$ quarks
   to the gravitational form factor of the nucleons. The cancellation of these contributions
   at $t=0$ shows that the gravitomagnetic form factor is close to zero for separate contributions of
  gluons and quarks, which supports the conjecture of \cite{Teryaev-s3,Teryaev:2006fk}.

\vspace{0.5cm}

 {\it The authors would like to thank
  M. Anselmino and E. Predazzi  for helpful discussion.
   The visit of O.S. to Torino was supported by the INFN-LTPh(JINR) agreement program.
This work was supported in part by Grants RFBR
06-02-16215 and RF MSE RNP 2.2.2.2.6546.
   }

\end{document}